\documentclass[prb,twocolumn,superscriptaddress,nobibnotes,amsmath,amssymb]{revtex4-1}
\usepackage{graphicx}         
\usepackage{bm}               
\usepackage{color}
\definecolor{darkblue}{rgb}{0.0,0.0,0.3}
\definecolor{goodblue}{rgb}{0.0,0.0,0.6}
\usepackage{float}
\usepackage{hyperref}
\hypersetup{
	colorlinks   = true, 
	urlcolor     = darkblue, 
	linkcolor    = darkblue, 
	citecolor   = darkblue 
}

\makeatletter
\def\vhrulefill#1{\leavevmode\leaders\hrule\@height#1\hfill \kern\z@}
\makeatother

\newenvironment{stretchtext}
{\par\setlength{\parfillskip}{0pt}}
{\par}

\begin{document}
	
\thispagestyle{empty}
\onecolumngrid

\begin{center}
	\textbf{\large Ballistic Majorana nanowire devices}
\end{center}
\begin{center}
\normalsize	{
			\"Onder~G\"ul,$^{1,*,\dagger,\ddagger}$ Hao~Zhang,$^{1,*,\dagger}$ Jouri~D.S.~Bommer,$^{1,*}$ Michiel~W.A.~de~Moor,$^{1}$ Diana~Car,$^{2}$ S\'ebastien~R.~Plissard,$^{2,\S}$
			Erik~P.A.M.~Bakkers,$^{1,2}$ Attila~Geresdi,$^{1}$ Kenji~Watanabe,$^{3}$ Takashi~Taniguchi,$^{3}$ Leo~P.~Kouwenhoven$^{1,4,\dagger}$
			}
\smallskip
\small

\emph{$^\mathit{1}$QuTech and Kavli Institute of Nanoscience, Delft University of Technology, 2600 GA Delft, The Netherlands}
	
\emph{$^\mathit{2}$Department of Applied Physics, Eindhoven University of Technology, 5600 MB Eindhoven, The Netherlands}
	
\emph{$^\mathit{3}$Advanced Materials Laboratory, National Institute for Materials Science, 1-1 Namiki, Tsukuba, 305-0044, Japan}
	
\emph{$^\mathit{4}$Microsoft Station Q Delft, 2600 GA Delft, The Netherlands}

\vspace*{0.5cm}

\end{center}

\twocolumngrid
\normalsize

\textbf{Majorana modes are zero-energy excitations of a topological superconductor that exhibit non-Abelian statistics \cite{1,2,3}. Following proposals for their detection in a semiconductor nanowire coupled to an s-wave superconductor \cite{4,5}, several tunneling experiments reported characteristic Majorana signatures \cite{6,7,9,12,13,14}. Reducing disorder has been a prime challenge for these experiments because disorder can mimic the zero-energy signatures of Majoranas \cite{15,16,17,18,19}, and renders the topological properties inaccessible \cite{20,21,22,23}. Here, we show characteristic Majorana signatures in InSb nanowire devices exhibiting clear ballistic transport properties. Application of a magnetic field and spatial control of carrier density using local gates generates a zero bias peak that is rigid over a large region in the parameter space of chemical potential, Zeeman energy, and tunnel barrier potential. The reduction of disorder allows us to resolve separate regions in the parameter space with and without a zero bias peak, indicating topologically distinct phases. These observations are consistent with the Majorana theory in a ballistic system \cite{24}, and exclude for the first time the known alternative explanations that invoke disorder \cite{15,16,17,18,19} or a nonuniform chemical potential \cite{25,26}.}

\begin{stretchtext}
Semiconductor nanowires are the primary contender for realizing a topological quantum bit (qubit) based on Majorana modes. Their confined geometry together with the highly tunable electronic properties readily allow for localizing Majoranas, engineering the coupling between Majoranas, and finally controlling the coupling between the topological superconductor and the external circuity. These requirements for the implementation of a Majorana qubit are challenging to achieve in other Majorana systems such as 2D and 3D topological insulators. Moreover, various basic networks \cite{28} and high-quality interfaces to different superconductors \cite{21,22,23} have \end{stretchtext}
\vspace{4.5pt}
\noindent \vhrulefill{0.25pt} \hspace*{7.2cm}\\

\footnotesize
\noindent $^*$ These authors contributed equally to this work.\\
$^{\dagger}$ Correspondence to \"O.G. (\href{mailto:onder_gul@g.harvard.edu}{onder\_gul@g.harvard.edu}) or H.Z. \\
\hspace*{7pt}(\href{mailto:h.zhang-3@tudelft.nl}{h.zhang-3@tudelft.nl}) or L.P.K. (\href{mailto:l.p.kouwenhoven@tudelft.nl}{l.p.kouwenhoven@tudelft.nl})\\
$^{\ddagger}$ Present address: Department of Physics, Harvard University,\\
\hspace*{7pt}Cambridge, MA 02138, USA\\
$^{\S}$ Present address: CNRS-Laboratoire d'Analyse et d'Architecture\\
\hspace*{7pt}des Syst\`emes (LAAS), Universit\'e de Toulouse, 7 avenue du\\
\hspace*{7pt}colonel Roche, F-31400 Toulouse, France

\normalsize
\noindent already been realized in semiconductor nanowires, fulfilling the further requirements for Majorana qubits. However, despite these advances in materials, alternative explanations have been proposed for the characteristic Majorana signatures. Most alternative explanations invoke bulk or interface disorder \cite{15,16,17,18,19} or a nonuniform chemical potential along the wire \cite{25,26}. Notable examples are weak antilocalization \cite{17}, Kondo effect \cite{18}, and Andreev levels \cite{19,25}, all shown to result in transport signatures mimicking those attributed to Majoranas. Here, we show characteristic Majorana signatures in nanowire devices that exhibit ballistic transport, ruling out all known disorder- or nonuniformity-based explanations for the first time.

\begin{figure*}
	\centering
	\includegraphics*[width=0.9\textwidth]{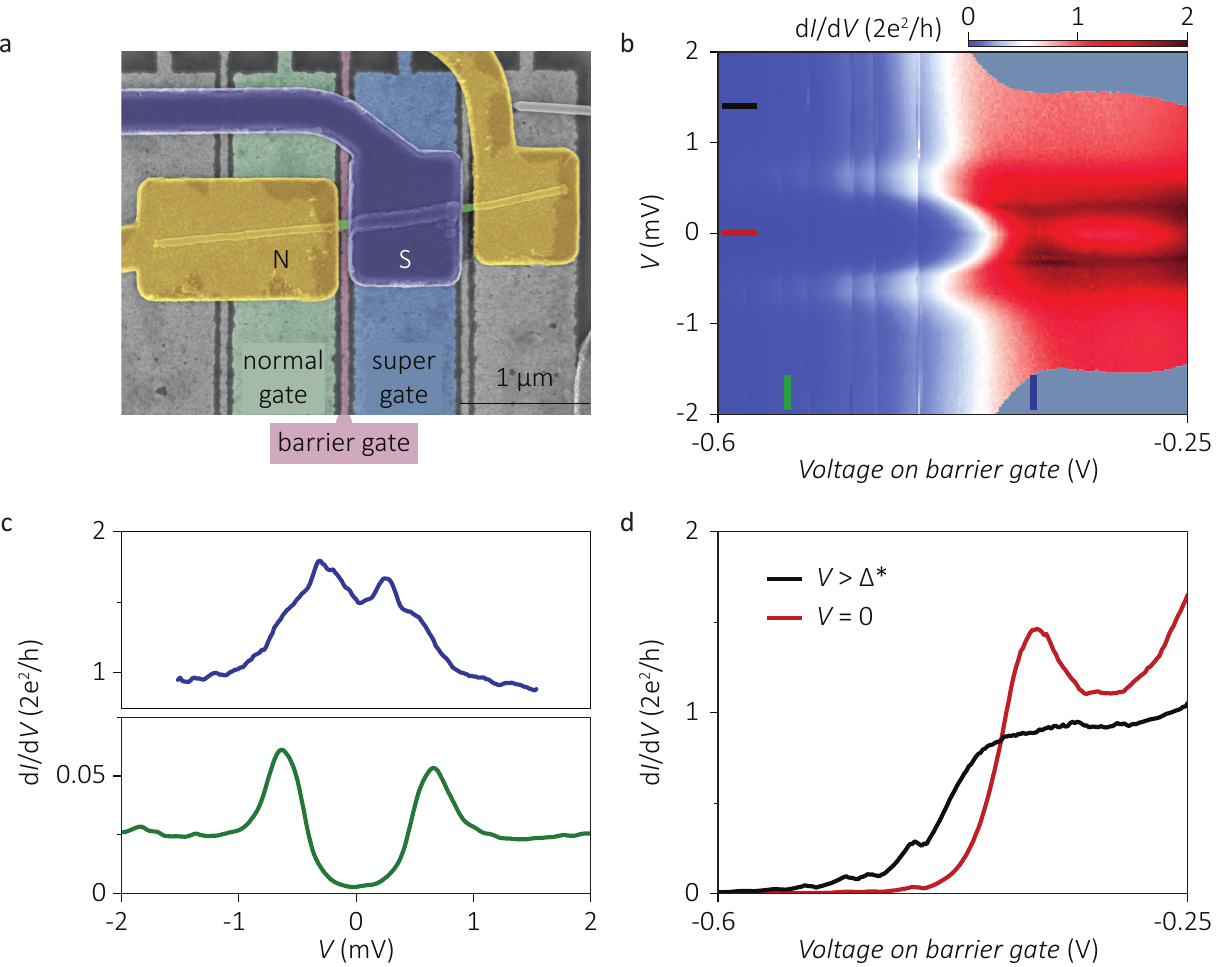}
	\caption{
		\textbf{Hybrid device and ballistic transport properties.} \textbf{a,} False-color electron micrograph of the measured device. The InSb nanowire (green) is contacted by a grounded NbTiN superconductor (purple) and two Au normal metal leads (yellow). The nanowire has a diameter of $\sim 80$\,nm. The local bottom gates (normal, barrier, and super gate) are separated from the nanowire by a boron nitride flake ($\sim 30$\,nm) and are operated individually. Two-terminal measurements are performed between N and S, while the other normal lead is floating. \textbf{b,} Differential conductance d$I$/d$V$ as a function of bias voltage $V$, and voltage on barrier gate (the other gate electrodes are grounded). Vertical lines at certain gate voltages are due to slow fluctuations in the electrostatic environment. \textbf{c,} Vertical line cuts from \textbf{b} at the gate voltages marked with colored bars. Top panel shows the d$I$/d$V$ from the transport regime in which the current is carried by a single fully-transmitting channel. We find an enhancement of conductance at small bias by more than a factor of 1.5 compared to the large-bias conductance of $2e^2/h$. Bottom panel is from the tunneling regime in which the current is carried by a single channel with low transmission. We extract an induced superconducting gap $\Delta^* = 0.65$\,meV. \textbf{d,} Horizontal line cuts from \textbf{b} at the bias voltages marked with colored bars. Subgap conductance ($V = 0$) shows an enhancement reaching $1.5 \times 2e^2/h$ when the large-bias conductance ($V = 1.4 \, \mathrm{mV} > \Delta^*$) has a quantized value of $2e^2/h$.
	}\label{bm1}
\end{figure*}

Figure~\ref{bm1}a shows the measured device consisting of an InSb nanowire (green) contacted with a grounded NbTiN superconductor (purple), and normal metal leads (yellow). The local bottom gate electrodes are separated from the nanowire by a boron nitride flake and are operated individually to allow for spatial control of the carrier density in the nanowire. We have realized our devices following our recently developed nanofabrication recipe which results in a high-quality InSb--NbTiN interface, a hard induced superconducting gap, and ballistic transport in the proximitized nanowire (see Ref.~\onlinecite{22} and \onlinecite{23}). All measurements are performed in a dilution refrigerator with an electron temperature of $\sim 50$\,mK. The data is taken by applying a bias voltage $V$ between the normal metal lead and the superconductor indicated by N and S, respectively, and monitoring the current flow. The other normal lead is kept floating.

Figure~\ref{bm1}b shows the differential conductance d$I$/d$V$ while varying $V$, and stepping the voltage applied to the barrier gate. Importantly, we find no signs of formation of quantum dots or any other localization effects. Vertical line cuts at the gate voltages indicated with colored bars are shown in Figure~\ref{bm1}c. Figure~\ref{bm1}c (bottom) is from the tunneling regime of the device where a sufficiently negative voltage on barrier gate locally depletes the noncovered nanowire section, and creates a tunnel barrier between the normal lead and the superconductor. In this regime we find an induced superconducting gap with a strong conductance suppression for subgap bias. The extracted gap value is $\Delta^* = 0.65$\,meV. Increasing the voltage on barrier gate first lowers the tunnel barrier and then removes it completely. Figure~\ref{bm1}c (top) is from the regime in which the noncovered nanowire section admits a single fully-transmitting transport channel. In this regime the subgap conductance is strongly enhanced due to Andreev reflection compared to the large-bias (above-gap) conductance of $2e^2/h$. The extracted enhancement factor $> 1.5$ implies a contact interface transparency $> 0.93$ \cite{23}. Figure~\ref{bm1}d shows the horizontal line cuts from Figure~\ref{bm1}b at the bias voltages indicated with colored bars. For a bias $V > \Delta^*$ we find a quantized conductance plateau at $2e^2/h$, a clear signature of a ballistic device. For zero bias voltage the strong Andreev enhancement is evident in the plateau region followed by a dip in conductance due to channel mixing \cite{23}. From the absence of quantum dots, the observed induced gap with a strongly reduced subgap density of states, high interface transparency, and quantized conductance, we conclude a very low disorder strength for our device, consistent with our recent findings \cite{23}.

\begin{figure*}
	\centering
	\includegraphics[width=0.891\textwidth]{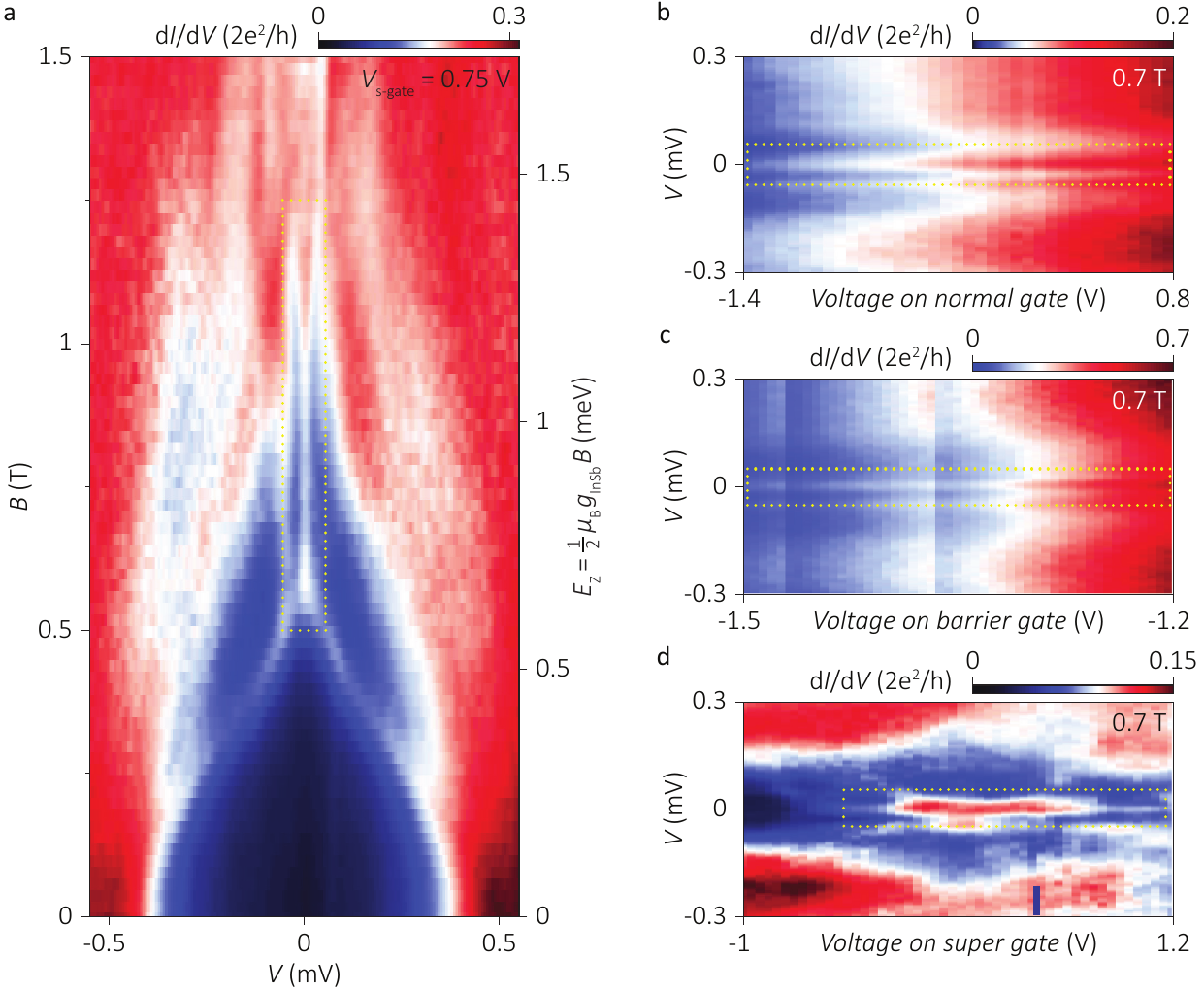}
	\caption{
		\textbf{Zero bias peak and its dependence on magnetic field and local gate voltages.} \textbf{a,} Differential conductance d$I$/d$V$ as a function of bias voltage $V$, and an external magnetic field $B$ along the nanowire axis for $V_\mathrm{s-gate}= 0.75$\,V. The magnetic field closes the induced gap at 0.55\,T and generates a zero bias peak which persists up to 1.2\,T. The right axis scales with Zeeman energy $E_z$ assuming a $g$ factor of 40 obtained independently \cite{30,31}. (Voltage on normal and barrier gate: 0\,V and $-1.4$\,V) \textbf{b,} d$I$/d$V$ as a function of $V$ and voltage on normal gate. The voltage on normal gate changes the conductance by more than a factor of 5 but does not affect the presence of the zero bias peak. \textbf{c,} d$I$/d$V$ as a function of $V$ and voltage on barrier gate. The voltage on barrier gate changes the conductance by nearly an order of magnitude but does not affect the presence of zero bias peak. \textbf{d,} d$I$/d$V$ as a function of $V$ and voltage on super gate. The zero bias peak persists for a finite gate voltage range. The blue bar indicates the voltage on super gate in \textbf{a}, \textbf{b} and \textbf{c}. Voltage on barrier gate is adjusted to keep the overall conductance the same when sweeping the voltage on super gate.
	}\label{bm2}
\end{figure*}

We now turn to the tunneling regime of the device where Majorana modes are characterized by a zero bias peak. To drive the nanowire device into the topological phase, we apply a magnetic field $B$ along the wire axis and tune the voltage applied to the super gate ($V_\mathrm{s-gate}$) which controls $\mu$, the chemical potential in the nanowire section underneath the superconductor. Figure~\ref{bm2}a shows that an increasing $B$ closes the induced gap at 0.55\,T and generates a zero bias peak rigidly bound to $V = 0$ up to 1.2\,T (line cuts in Suppl. Figure~1a). The gap closure is expected to occur for a Zeeman energy $E_z \gtrsim \Delta^*$. From linear interpolation we find $g \gtrsim 40$ which matches our independent measurements \cite{30,31}. Converting the $B$ axis into a Zeeman energy $E_z$ scale (right vertical axis), we find that the zero bias peak is bound to zero over 0.75\,meV, a range in Zeeman energy that is 30 times larger than the peak width (the full width at half maximum, $\mathrm{FWHM} \sim 20$\,$\mu$eV, see Suppl. Figure~1c and Suppl. Figure~4). This excludes a level crossing as the origin for our zero bias peak \cite{18}. We note that all our devices show a significant increase of subgap density of states for the magnetic fields required for topological phase transition. This behavior is likely due to vortex formation or a short mean free path \cite{32,33} in our NbTiN film, and is a subject of our future studies. The formation of vortices is speculated to create a dissipation channel \cite{24}, the leading hypothetical mechanism that limits our zero bias peak height from reaching the quantized value of $2e^2/h$. An unambiguous observation of a zero bias peak quantization remains an outstanding challenge for Majorana experiments in semiconductor nanowires.

\begin{figure*}[t!]
	\centering
	\includegraphics[width=\textwidth]{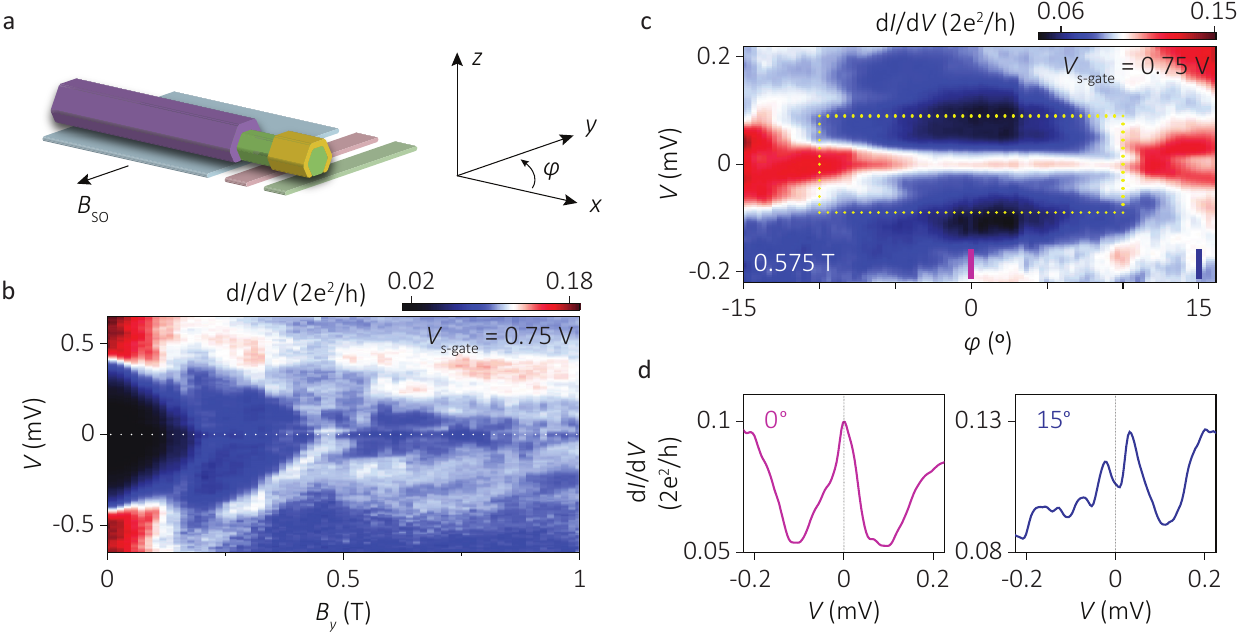}
	\caption{
		\textbf{Dependence of zero bias peak on magnetic field orientation.} \textbf{a,} Orientation of the nanowire device. The wire is along $x$ and the spin--orbit field $B_\mathrm{so}$ is along $y$. The substrate plane is spanned by $x$ and $y$. $\varphi$ is the angle between the $x$ axis and the orientation of the external magnetic field in the plane of the substrate. \textbf{b,} Differential conductance d$I$/d$V$ as a function of bias voltage $V$, and an external magnetic field along the $y$ axis. Application of a magnetic field along $B_\mathrm{so}$ closes the induced gap but does not generate a zero bias peak. \textbf{c,} d$I$/d$V$ as a function of $V$, and in-plane rotation of the magnetic field with a magnitude of 0.575\,T. The zero bias peak appears in an angle range in which the external magnetic field is mostly aligned with the wire. We attribute the low conductance region around the zero bias peak to the induced gap. Orienting the magnetic field away from the wire axis and more towards $B_\mathrm{so}$ closes the induced gap and splits the zero bias peak (see line cuts in \textbf{d}). \textbf{d,} Vertical line cuts from \textbf{c} at the angles indicated with colored bars. For $\varphi = 0^{\circ}$ the zero bias peak is present, which is split for $\varphi = -15^{\circ}$.
	}\label{bm3}
\end{figure*}

The origin of zero bias peak can be spatially resolved by varying the voltages applied to individual gates. Figure~\ref{bm2}b shows that the presence of the zero bias peak is not affected when gating the wire section underneath the normal contact which changes the conductance by more than a factor of 5 (see also Suppl. Figure~1d). Extending the same analysis to the noncovered wire section yields the same result (Figure~\ref{bm2}c), i.e., changing the tunnel barrier conductance by nearly an order of magnitude does not split the zero bias peak, nor makes it disappear (see also Suppl. Figure~1e). In contrast, Figure~\ref{bm2}d shows that the zero bias peak is present over a finite range in voltage applied to the super gate (line cuts in Suppl. Figure~1f). This indicates that proper tuning of $\mu$ is essential for the appearance of the zero bias peak. The observation of a zero bias peak that does not split when changing the tunnel barrier conductance (Figure~\ref{bm2}c) excludes Kondo effect \cite{18} and crossing of Andreev levels \cite{19} as the origin of our zero bias peak. Most importantly, it rules out an explanation provided by recent theory work \cite{25} demonstrating trivial Andreev levels localized near the noncovered wire section that are bound to zero energy for varying $E_z$ and $\mu$, but quickly split to finite energies for varying tunnel barrier strength. Here we demonstrate for the first time a zero bias peak rigidly bound to $V = 0$ over a changing tunnel barrier conductance---a behavior observed in all devices (Suppl. Figure~5-7). From the combined analysis (Figure~\ref{bm2}b-d) we conclude that the zero bias peak originates in the wire section underneath the superconductor, consistent with a Majorana interpretation.

\begin{stretchtext}
In a Majorana nanowire \cite{4,5}, the existence of a topological phase strictly requires an external magnetic field with a finite component perpendicular to the spin--orbit field $B_\mathrm{so}$, see Figure~\ref{bm3}a. An external field along the wire fulfills this requirement, shown in Figure~\ref{bm2}a. In contrast, Figure~\ref{bm3}b shows that an external magnetic field parallel to $B_\mathrm{so}$ does not generate a zero bias peak for the same gate settings in Figure~\ref{bm2}a. Figure~\ref{bm3}c shows the dependence of the zero bias peak on the direction of the external field. The zero bias peak is limited to an angle range where the external field is mostly aligned with the wire, perpendicular to $B_\mathrm{so}$ (see Suppl. Figure~2 for a measurement in a larger angle range). We observe a low conductance region around the zero bias peak, indicating the induced gap. Orienting the magnetic field away from the wire axis and more towards $B_\mathrm{so}$ closes the induced gap and splits the zero bias peak. This is indicated by the vertical line cuts from Figure~\ref{bm3}c at marked angles, shown in Figure~\ref{bm3}d. A gap closing is expected for the critical an-
\end{stretchtext}

\begin{figure*}[p!]
\centering
\includegraphics[width=0.83\textwidth]{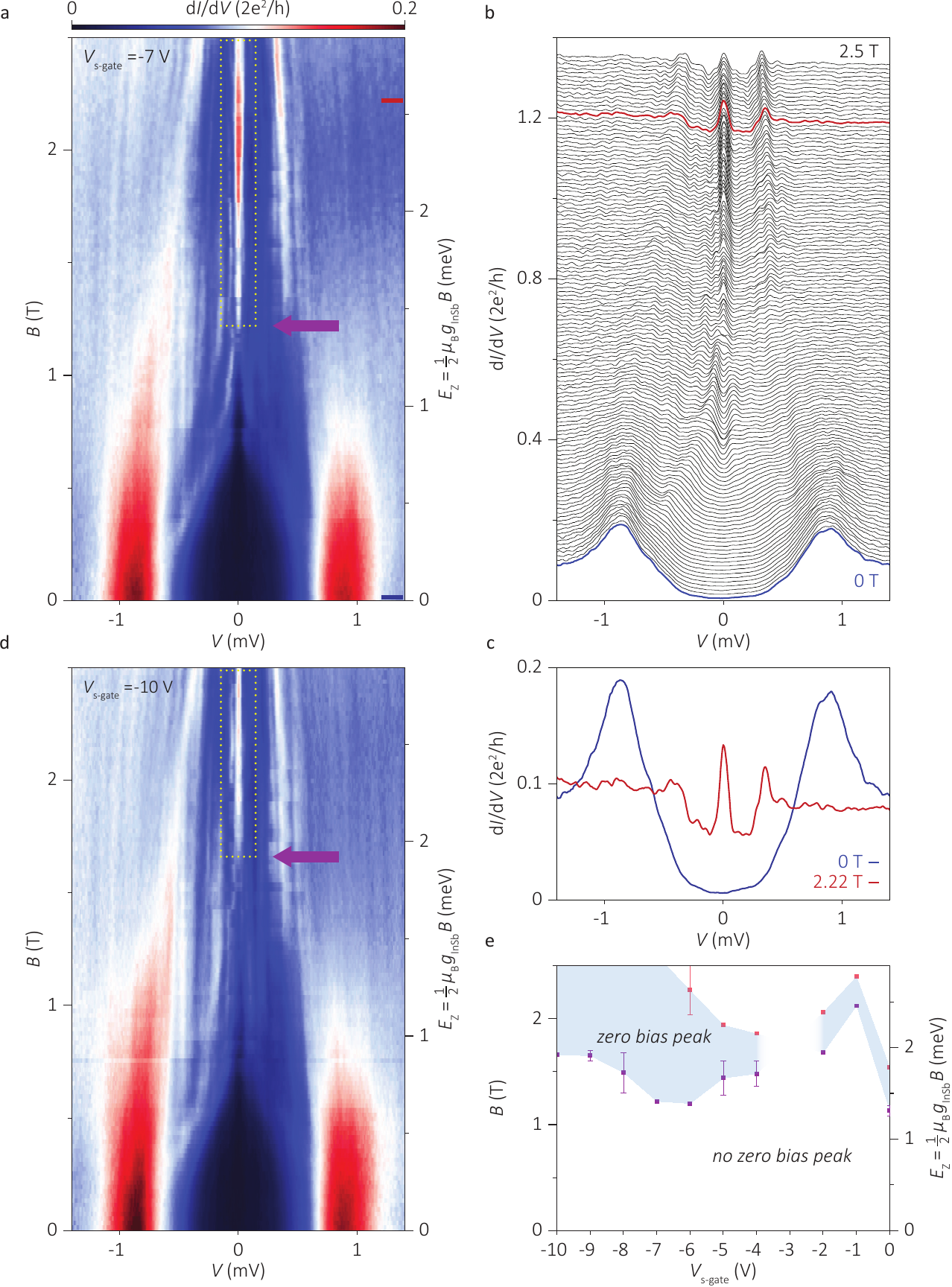}
\caption{
	\textbf{Zero bias peak and phase diagram.} \textbf{a,} Differential conductance d$I$/d$V$ of another device measured as a function of bias voltage $V$, and an external magnetic field $B$ along the nanowire axis. We find an induced gap $\Delta^*=0.9$\,meV at zero magnetic field. Increasing the magnetic field closes the induced gap at $\sim$1\,T and generates a zero bias peak that persists up to at least 2.5\,T. The right axis scales with Zeeman energy $E_z$ assuming $g_\mathrm{InSb}=40$ obtained independently \cite{30,31}. The purple arrow at 1.22\,T indicates the onset of the zero bias peak. \textbf{b,} Line cuts from \textbf{a} with $0.01 \times 2e^2/h$ offset. \textbf{c,} Line cuts from \textbf{a} and \textbf{b} at 0 and 2.22\,T. \textbf{d,} Same as \textbf{a} but with a different super gate voltage $V_\mathrm{s-gate}=-10$\,V. For this super gate voltage, the onset of the zero bias peak is at a larger magnetic field of 1.66\,T, as marked by a purple arrow. \textbf{e,} Phase diagram constructed by the onset and the end of the zero bias peak in magnetic field for different super gate voltages. The purple squares denote the onset, pink the end. For $V_\mathrm{s-gate}=-3$\,V no zero bias peak is observed.
}\label{bm4}
\end{figure*}

\noindent gle $\varphi_c$ given by the projection rule \cite{34,35} $E_z \: \mathrm{sin}(\varphi_c) = \Delta^*$. From the observed gap $\Delta^* = 175$\,$\mu$eV at $B = 0.575$\,T and a $g$ factor of 40, we obtain $\varphi_c = 15^{\circ}$, agreeing well with the observed value of $\varphi_c \sim 10^{\circ}$ (a reduction in $\varphi_c$ is expected due to orbital effect of the external magnetic field \cite{36}). Finally, in Suppl. Figure~2 we show that increasing $B$ decreases $\varphi_c$, a behavior consistent with the projection rule.

We now turn our attention to an identical device but with a longer proximitized wire section (1.2\,$\mu$m, see Suppl. Figure~3a). Figure~\ref{bm4}a-c show an induced gap $\Delta^* = 0.9$\,meV at zero magnetic field, significantly larger than the device in Figure~\ref{bm1}-\ref{bm3}. As a result, the induced gap closes at a higher magnetic field ($\sim 1$\,T). The zero bias peak is visible and unsplit over a range of at least 1.3\,T, corresponding to a Zeeman energy scale $> 1.5$\,meV. The FWHM is around 0.07\,meV yielding a ratio ZBP-range$/\mathrm{FWHM} \gtrsim 20$ (Suppl. Figure~4). A disorder-free Majorana theory model with parameters extracted from this device (geometry, induced gap, spin--orbit coupling, temperature) finds perfect agreement between simulation \cite{24} and our data (Figure~\ref{bm4}a). Suppl. Figure~3b and c shows that the zero bias peak position is robust against a change in conductance when varying the voltage applied to the normal and the barrier gate, ruling out the trivial Andreev-level explanation \cite{25} consistent with our earlier discussion (Figure~\ref{bm2}b and c). In contrast to normal and barrier gate, the voltage applied to the super gate changes the onset and the end of the zero bias peak in magnetic field. Figure~\ref{bm4}d shows that for $V_\mathrm{s-gate} = - 10$\,V the zero bias peak appears at a higher magnetic field compared to Figure~\ref{bm4}a where $V_\mathrm{s-gate} = - 7$\,V (1.66\,T vs. 1.22\,T). We have extended this analysis for $-10 \, \mathrm{V} \leq V_\mathrm{s-gate}\leq 0\, \mathrm{V}$ and marked the magnetic field values at which the zero bias peak starts and ends (Suppl. Figure~3d). The resulting phase diagram is shown in Figure~\ref{bm4}e. For large negative voltages applied to the super gate, we find a region in which the zero bias peak persists for large ranges of magnetic field and $V_\mathrm{s-gate}$, indicating the topological phase. We attribute the appearance of a trivial phase at large magnetic fields above the topological phase to multi-channel occupation in the proximitized wire section \cite{24,25}. A precise knowledge of the phase boundaries requires theory including finite-size effects \cite{37}, the orbital effect of the magnetic field \cite{36}, and an accurate electrostatic modeling of the device \cite{38}, and will be addressed in future studies.

In conclusion, the presented experiments demonstrate zero bias peaks over an extended range in Zeeman energy and gate voltage in devices that show clear ballistic transport properties, and reveal the distinct phases in the topology of Majorana wires. These observations exclude all known alternative explanations for our zero bias peaks that are based on disorder.

\vspace*{2\baselineskip}
{\centering	\textbf{Methods}\par}
\vspace*{1\baselineskip}

\noindent \textbf{Nanowire growth and device fabrication.} InSb nanowires have been grown by Au-cata\-lyzed Vapor-Liquid-Solid mechanism in a Metal Organic Vapor Phase Epitaxy reactor. The InSb nanowires are zinc blende, along [111] crystal direction, and are free of stacking faults and dislocations \cite{39}. As-grown nanowires are deposited one-by-one using a micro-manipulator \cite{40} on a substrate patterned with local gates covered by a $\sim 30$\,nm thick hBN dielectric. The contact deposition process starts with resist development followed by oxygen plasma cleaning. Then, the chip is immersed in a sulfur-rich ammonium sulfide solution diluted by water (with a ratio of 1:200) at $60^{\circ}$C for half an hour \cite{41}. At all stages care is taken to expose the solution to air as little as possible. For normal metal contacts \cite{30,31}, the chip is placed into an evaporator. A 30 second helium ion milling is performed in situ before evaporation of Cr/Au (10\,nm/125\,nm) at a base pressure $< 10^{-7}$\,mbar. For superconducting contacts \cite{22,23}, the chip is mounted in a sputtering system. After 5 seconds of in situ argon plasma etching at a power of 25\,Watts and a pressure of 10\,mTorr, 5\,nm NbTi is sputtered followed by 85\,nm NbTiN.

\hypersetup{urlcolor=goodblue}
\medskip
\noindent \textbf{Data availability.} All data are available at \href{http://doi.org/10.4121/uuid:b3f993a7-1b8b-4fd8-8142-5fa577027cdd}{doi.org/\hspace{1pt}10.4121/uuid:b3f993a7-1b8b-4fd8-8142-5fa577027cdd} (Ref.~\onlinecite{data}).

\vspace*{2.2\baselineskip}
{\centering	\textbf{Acknowledgments}\par}
\vspace*{1\baselineskip}

We thank A.R. Akhmerov, O.W.B. Benningshof, M.C. Cassidy, S. Goswami, J. Kammhuber, V. Mourik, M. Quintero-P\'erez, J. Shen, M. Wimmer, D.J. van Woerkom, and K. Zuo for discussions and assistance. This work has been supported by the Netherlands Organisation for Scientific Research (NWO), European Research Council (ERC), and Microsoft Corporation Station Q.

\vspace*{2.65\baselineskip}
{\centering	\textbf{Author contributions}\par}
\vspace*{1\baselineskip}

\"O.G., H.Z., and J.D.S.B fabricated the devices, performed the measurements, and analyzed the data. M.W.A.d.M. contributed to the device fabrication. D.C., S.R.P. and E.P.A.M.B. grew the InSb nanowires. A.G. contributed to the experiments. K.W. and T.T. synthesized the hBN crystals. L.P.K. supervised the project. \"O.G., H.Z., and J.D.S.B. co-wrote the paper. All authors commented on the manuscript.

\hypersetup{urlcolor=darkblue}
\clearpage
\onecolumngrid
\setlength\parindent{0pt}
\linespread{1.2}
\renewcommand*{\citenumfont}[1]{S#1}
\renewcommand*{\bibnumfmt}[1]{[S#1]}

\begin{center}
	\textbf{\large Supplementary Information: Ballistic Majorana nanowire devices}
\end{center}
\begin{center}
	\normalsize	{
		\"Onder~G\"ul,$^{*,\dagger}$ Hao~Zhang,$^{*,\dagger}$ Jouri~D.S.~Bommer,$^{*}$ Michiel~W.A.~de~Moor, Diana~Car, S\'ebastien~R.~Plissard,
		Erik~P.A.M.~Bakkers, Attila~Geresdi, Kenji~Watanabe, Takashi~Taniguchi, Leo~P.~Kouwenhoven$^{\dagger}$
	}
	\bigskip
	\small
	
	$^*$ These authors contributed equally to this work.
	
	$^{\dagger}$ Correspondence to \"O.G. (\href{mailto:onder_gul@g.harvard.edu}{onder\_gul@g.harvard.edu}) or H.Z. (\href{mailto:h.zhang-3@tudelft.nl}{h.zhang-3@tudelft.nl}) or L.P.K. (\href{mailto:l.p.kouwenhoven@tudelft.nl}{l.p.kouwenhoven@tudelft.nl})
	
	\vspace*{5cm}
	
\end{center}

\normalsize

\begin{center}
\textbf{List of supplementary figures}
\end{center}

\bigskip

\textbf{Supplementary Figure 1 $|$} Line cuts from main text Figure 2.

\medskip

\textbf{Supplementary Figure 2 $|$} Dependence of the zero bias peak on the orientation of an in-plane magnetic field.

\medskip

\textbf{Supplementary Figure 3 $|$} Zero bias peak in a large range of magnetic field and local gate voltages.

\medskip

\textbf{Supplementary Figure 4 $|$}  Zero bias peak height and width.

\medskip

\textbf{Supplementary Figure 5 $|$} Additional device 1 - ballistic transport properties.

\medskip

\textbf{Supplementary Figure 6 $|$} Additional device 1 - zero bias peak in a large range of magnetic field and local \\\hspace*{4.52cm} gate voltages.

\medskip

\textbf{Supplementary Figure 7 $|$} Additional device 2 - zero bias peak in a large range of magnetic field and local \\\hspace*{4.52cm} gate voltages.

\clearpage
\begin{figure*}[h!]
\centering
\includegraphics[height=1\textheight]{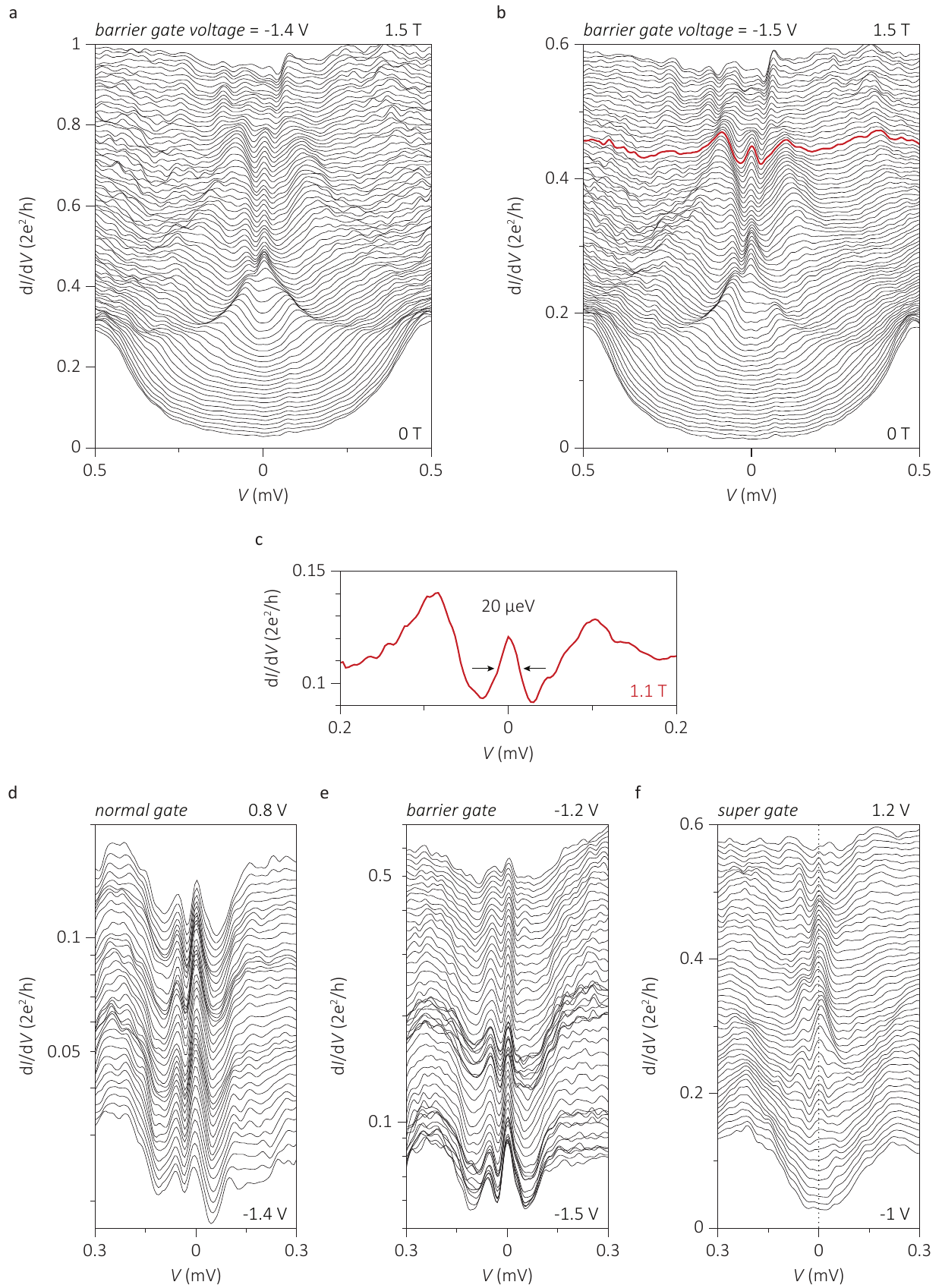}
\label{bms1}
\end{figure*}
\clearpage
Supplementary Figure 1 $|$ \textbf{Line cuts from main text Figure 2.}
\textbf{a,} Line cuts from main text Figure~\ref{bm2}a with $0.01 \times 2e^2/h$ offset. An increasing magnetic field closes the induced gap at 0.55\,T and generates a zero bias peak up to 1.2\,T. \textbf{b,} Same as \textbf{a} but for a larger tunnel barrier strength. Offset: $0.006 \times 2e^2/h$. \textbf{c,} Line cut from 1.1\,T. We find the full width at half maximum of the zero bias peak to be 20\,$\mu$eV. \textbf{d,} Line cuts from main text Figure~\ref{bm2}b in logarithmic scale (without offset). A variation in voltage on the normal gate ($-1.4 \, \mathrm{V} \leq V_\mathrm{n-gate} \leq 0.8$\,V) changes the conductance by more than a factor of 5, but does not remove the zero bias peak. \textbf{e,} Line cuts from main text Figure~\ref{bm2}c in logarithmic scale (without offset). A variation in voltage on the barrier gate ($-1.5 \, \mathrm{V} \leq V_\mathrm{b-gate} \leq -1.2$\,V) changes the conductance by nearly an order of magnitude, but does not remove the zero bias peak. \textbf{f,} Line cuts from main text Figure~\ref{bm2}d with $0.006 \times 2e^2/h$ offset. A variation in voltage on the super gate ($-1 \mathrm{V} \leq V_\mathrm{s-gate} \leq 1.2$\,V) affects the presence of the zero bias peak, which is stable for a finite gate voltage range.

\clearpage
\begin{figure*}
\centering
\includegraphics[width=1\columnwidth]{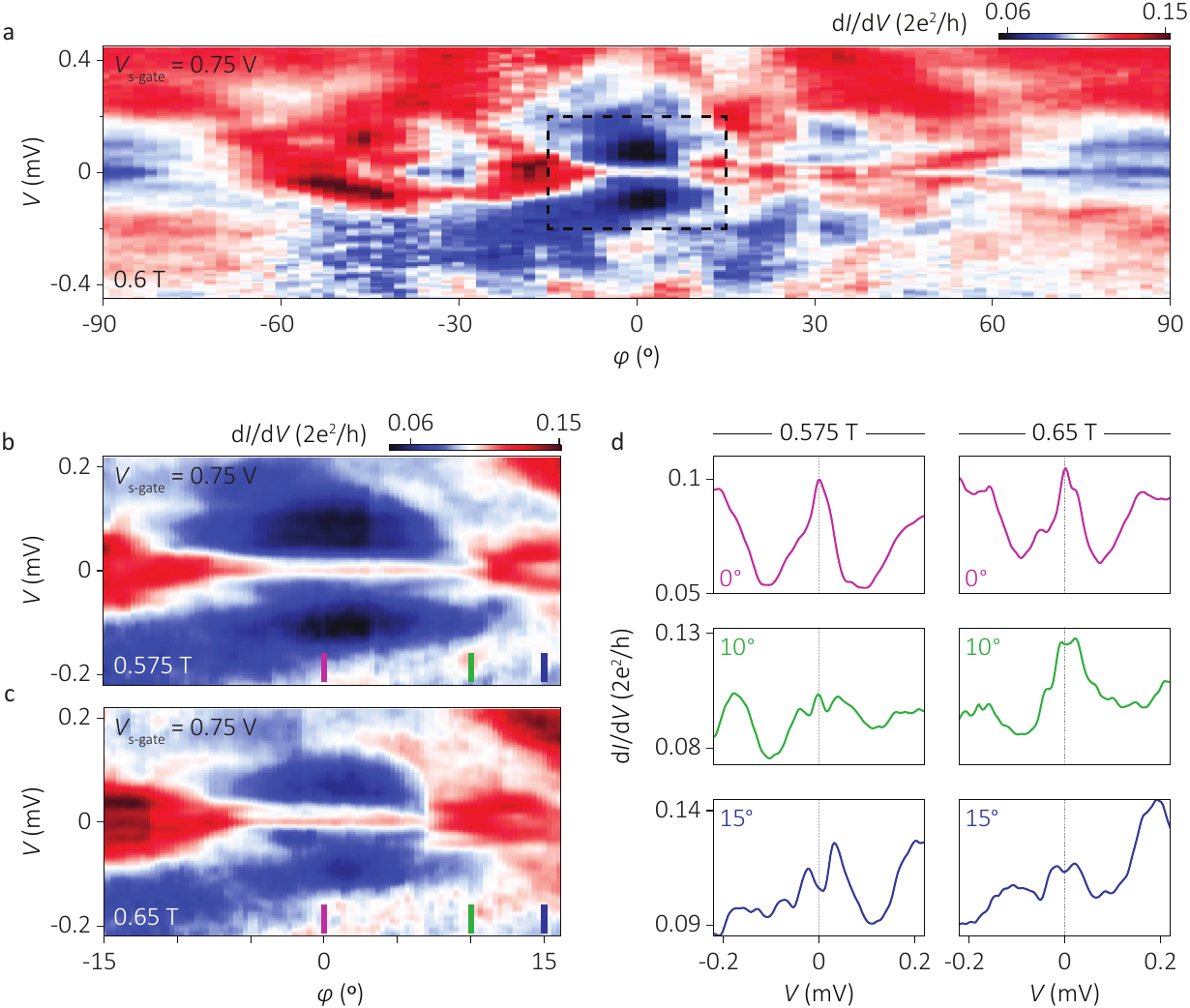}
\label{bms2}
\end{figure*}
Supplementary Figure 2 $|$ \textbf{Dependence of the zero bias peak on the orientation of an in-plane magnetic field.}
\textbf{a,} Differential conductance d$I$/d$V$ as a function of bias voltage $V$, and in-plane rotation of the magnetic field with a magnitude of 0.6\,T. $\varphi = 0^{\circ}$ corresponds to an external field along the wire, $\varphi = \pm 90^{\circ}$ to an external field parallel to the spin--orbit field $B_\mathrm{so}$. The zero bias peak is present in an angle range ($|\varphi| < 10^{\circ}$) when the external magnetic field is mostly aligned with the wire. We observe a low conductance region around the zero bias peak, indicating the induced gap. Orienting the magnetic field away from the wire axis and more towards $B_\mathrm{so}$ closes the induced gap and splits the zero bias peak. We do not observe a stable zero bias peak for $|\varphi| > 10^{\circ}$ in the entire angle range. The dashed square indicates the angle range shown in main text Figure~\ref{bm3}c. \textbf{b,} \textbf{c,} d$I$/d$V$ as a function of $V$, and in-plane rotation of the magnetic field with two different magnitudes. Increasing the magnetic field decreases the angle range of the zero bias peak (compare \textbf{b} and \textbf{c}). \textbf{d,} Vertical line cuts from \textbf{b} and \textbf{c} at the angles indicated with colored bars. Top panels: For $\varphi = 0^{\circ}$ the zero bias peak is present for both magnetic field magnitudes. Bottom panels: For $\varphi = 15^{\circ}$ no zero bias peak is present for both magnitudes. Middle panels: For $\varphi = 10^{\circ}$ the zero bias peak is present only for 0.575\,T, while is split for 0.65\,T.

\clearpage
\begin{figure*}
\centering
\includegraphics[height=1\textheight]{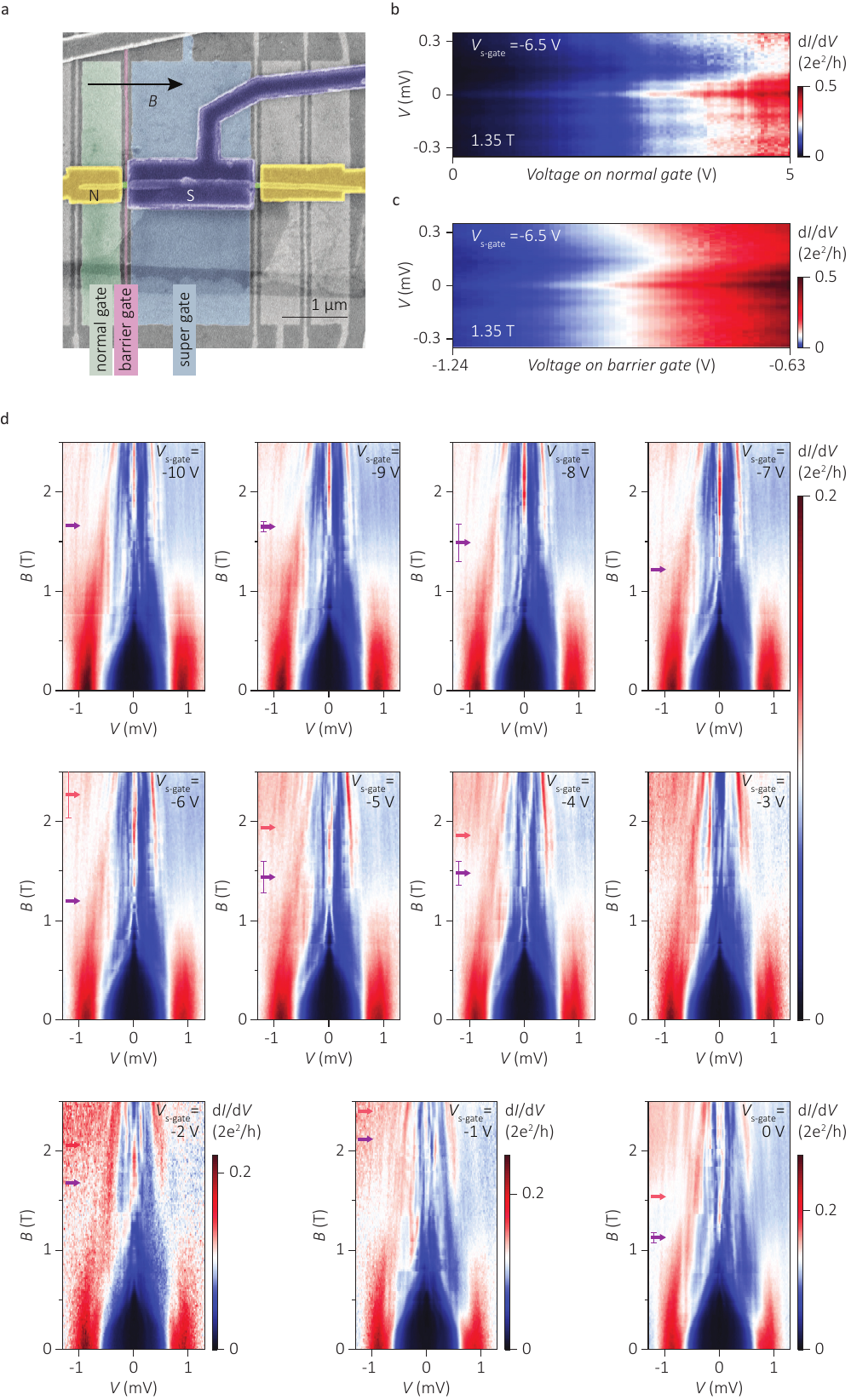}
\label{bms3}
\end{figure*}
\clearpage
Supplementary Figure 3 $|$ \textbf{Zero bias peak in a large range of magnetic field and local gate voltages.}
\textbf{a,} False-color electron micrograph of the measured device. \textbf{b,} \textbf{c,} Differential conductance d$I$/d$V$ as a function of bias voltage $V$, and voltages on normal and barrier gate, respectively. A variation in each gate voltage changes the conductance by an order of magnitude, but does not affect the presence of the zero bias peak. \textbf{d,} d$I$/d$V$ as a function of $V$ and an external magnetic field $B$ along the nanowire axis, measured at different voltages on super gate ($-10 \, \mathrm{V} \leq V_\mathrm{s-gate} \leq 0$\,V). A variation in $V_\mathrm{s-gate}$ changes the magnetic field range in which the zero bias peak appears. The purple (pink) arrows indicate the onset (end) of the zero bias peak in external magnetic field. When changing the super gate voltage, we adjust the tunnel gate voltage to keep the overall conductance the same.

\clearpage
\begin{figure*}
\centering
\vspace*{5cm}
\includegraphics[width=0.9\columnwidth]{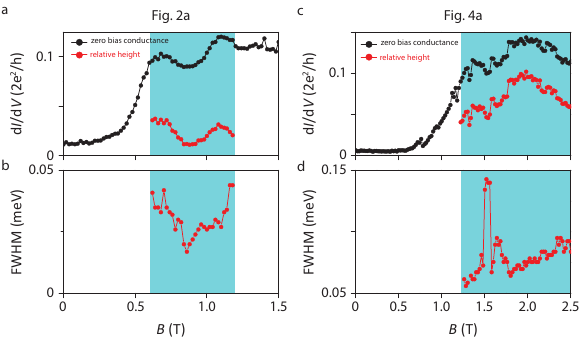}
\label{bms4}
\end{figure*}
Supplementary Figure 4 $|$ \textbf{Zero bias peak height and width.}
\textbf{a,} The absolute and the relative height of the zero bias peak extracted from main text Figure~\ref{bm2}a. The difference between the absolute and the relative height is the subgap conductance around zero bias for each magnetic field. \textbf{b,} The full width at half maximum (FWHM) of the zero bias peak extracted from main text Figure~\ref{bm2}a. \textbf{c, d,} Same as \textbf{a} and \textbf{b} but for the zero bias peak from main text Figure~\ref{bm4}a.

\clearpage
\begin{figure*}
\centering
\vspace*{2cm}
\includegraphics[width=0.9\columnwidth]{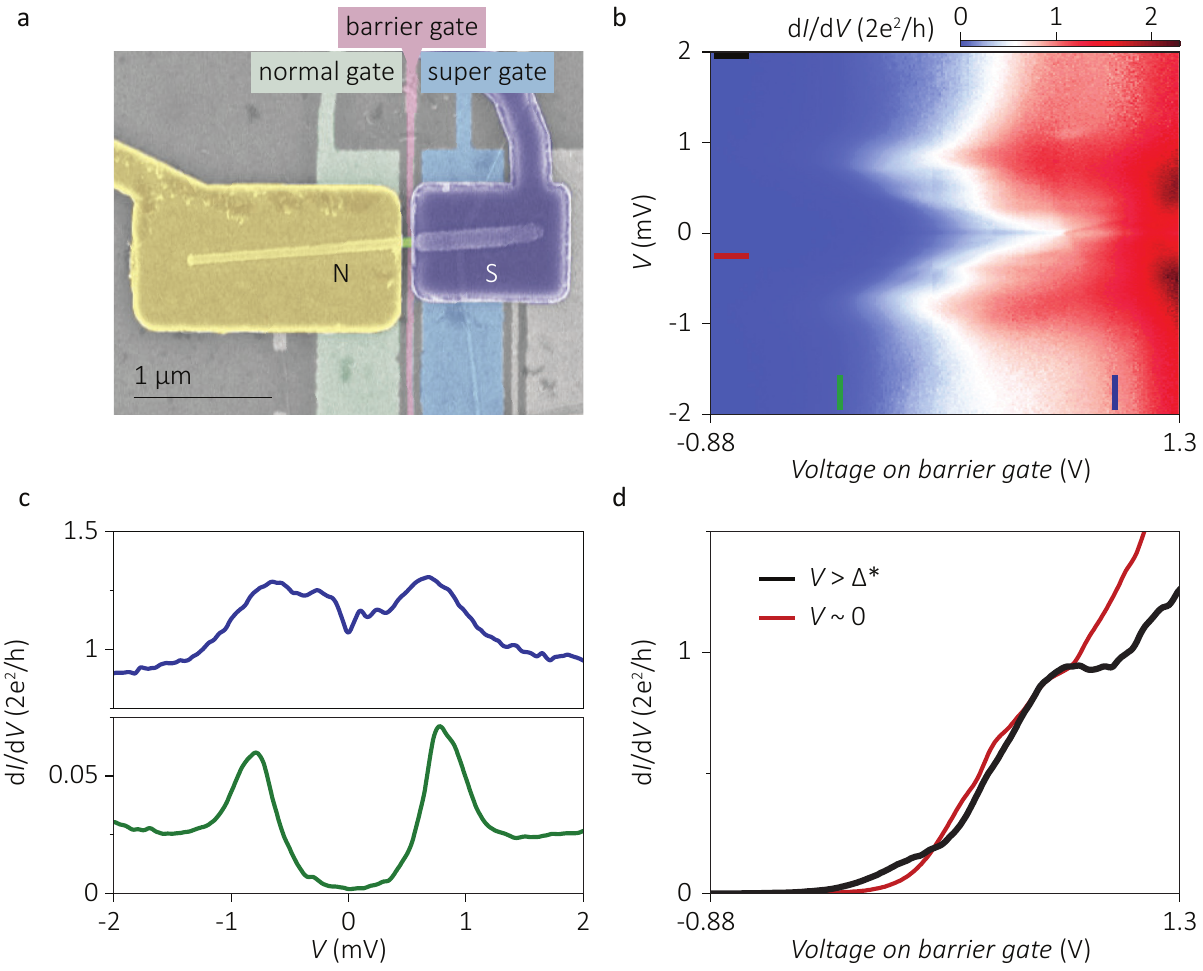}
\label{bms5}
\end{figure*}
Supplementary Figure 5 $|$ \textbf{Additional device 1 - ballistic transport properties.}
\textbf{a,} False-color electron micrograph of the measured device. \textbf{b,} Differential conductance d$I$/d$V$ as a function of bias voltage $V$, and voltage on barrier gate. \textbf{c,} Vertical line cuts from \textbf{b} at the gate voltages marked with colored bars. Top panel shows the d$I$/d$V$ from the transport regime in which the current is carried by a single fully-transmitting channel. We find an enhancement of conductance at small bias by a factor of 1.25 compared to the large-bias conductance of $2e^2/h$. Bottom panel is from the tunneling regime. We extract an induced superconducting gap $\Delta^* = 0.75$\,meV. \textbf{d,} Horizontal line cuts from \textbf{b} at the bias voltages marked with colored bars. Large-bias conductance ($V = 2 \, \mathrm{mV} > \Delta^*$) shows a quantized plateau of $2e^2/h$. The subgap conductance ($V = -0.25$\,mV) is enhanced within the gate voltage range in which the large-bias conductance is quantized. We excluded a series resistance of 0.5\,k$\Omega$, solely to account for the contact resistance of the normal lead\cite{s1,s2}.

\clearpage
\begin{figure*}[h!]
\centering
\includegraphics[width=1\columnwidth]{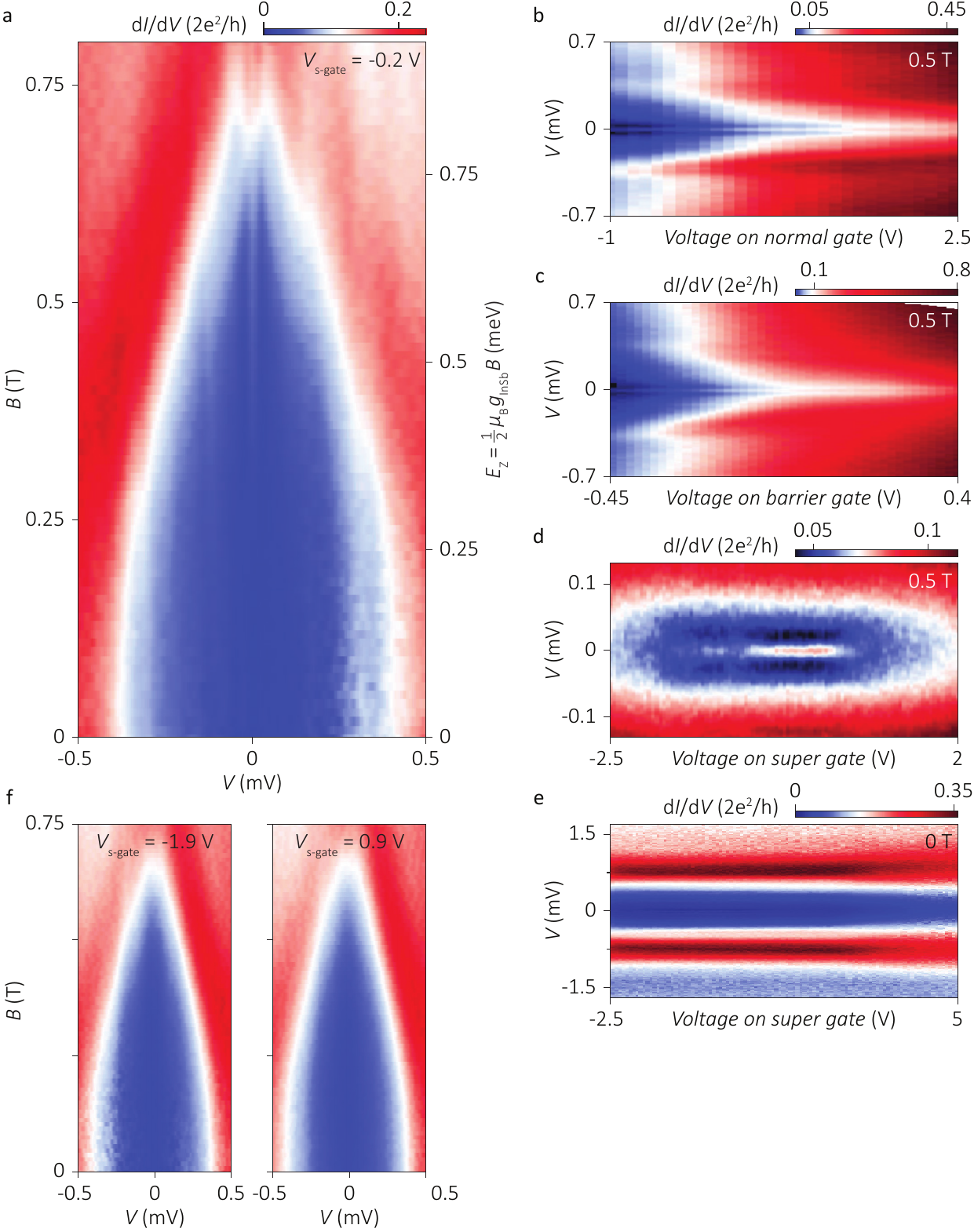}
\label{bms6}
\end{figure*}

\clearpage

Supplementary Figure 6 $|$ \textbf{Additional device 1 - zero bias peak in a large range of magnetic field and local gate voltages.}
\textbf{a,} Differential conductance d$I$/d$V$ as a function of bias voltage $V$, and an external magnetic field $B$ along the nanowire axis. Application of a magnetic field generates a zero bias peak at 0.3\,T. The zero bias peak persists up to 0.7\,T and splits for larger magnetic fields. The right axis scales with Zeeman energy $E_z$ assuming a $g$ factor of 40. \textbf{b,} d$I$/d$V$ as a function of $V$ and voltage on the normal gate. The voltage on the normal gate changes the conductance by a factor of 10 but does not affect the presence of the zero bias peak. \textbf{c,} d$I$/d$V$ as a function of $V$ and voltage on the barrier gate. The voltage on the barrier gate changes the conductance by a factor of 20 but does not affect the presence of the zero bias peak. \textbf{d,} d$I$/d$V$ as a function of $V$ and voltage on the super gate. The zero bias peak is stable for a finite range of voltages on the super gate. \textbf{e,} Same as \textbf{d} but at zero magnetic field. No zero bias peak is present. \textbf{f,} Same as \textbf{a} but for different voltages on the super gate ($V_\mathrm{s-gate}$). No zero bias peak is present for the measured magnetic field range for $V_\mathrm{s-gate} = -1.9$\,V and $V_\mathrm{s-gate} = 0.9$\,V, consistent with \textbf{d}.

\clearpage
\begin{figure}
\centering
\includegraphics[width=1\columnwidth]{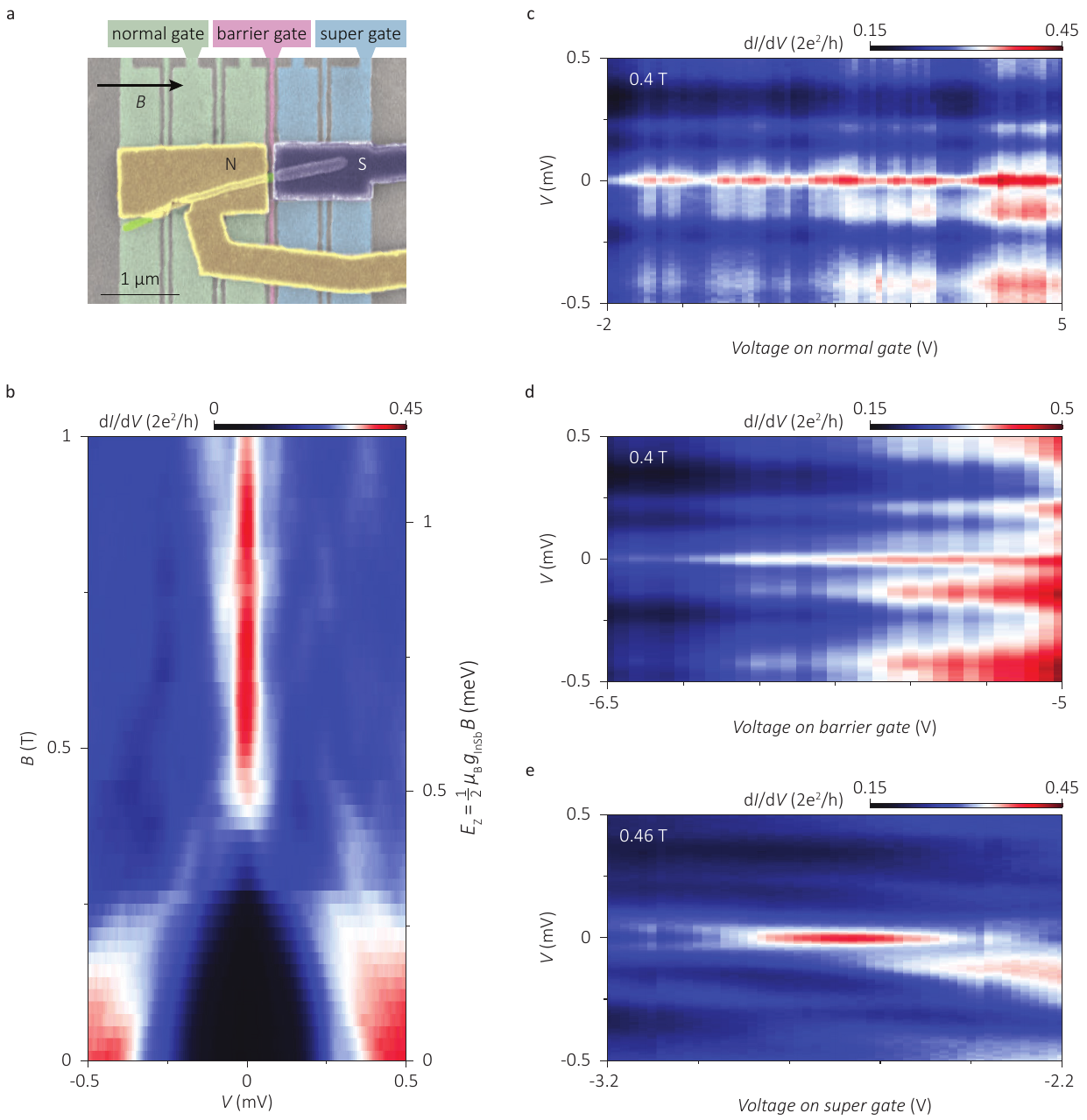}
\label{bms7}
\end{figure}
\vspace*{-30pt}
Supplementary Figure 7 $|$ \textbf{Additional device 2 - zero bias peak in a large range of magnetic field and local gate voltages.}
\textbf{a,} False-color electron micrograph of the measured device. \textbf{b,} Differential conductance d$I$/d$V$ as a function of bias voltage $V$ and magnetic field $B$. Increasing the magnetic field closes the gap and generates a zero bias peak which persists up to at least 1\,T. The right axis scales with Zeeman energy assuming $g_\mathrm{InSb} = 40$. Gate voltages are fixed at $V_\mathrm{n-gate} = 1$\,V, $V_\mathrm{b-gate} = -5.5$\,V, and $V_\mathrm{s-gate} = -2.8$\,V for normal, barrier, and super gate, respectively. \textbf{c,} d$I$/d$V$ as a function of $V$, and normal gate voltage $V_\mathrm{n-gate}$. A large variation in $V_\mathrm{n-gate}$ can modulate the overall conductance but it does not remove the zero bias peak. Taken at $B = 0.4$\,T, $V_\mathrm{b-gate} = -5.5$\,V, and $V_\mathrm{s-gate} = -2.85$\,V. \textbf{d,} d$I$/d$V$ as a function of $V$, and barrier gate voltage. Changing the tunnel barrier conductance by more than a factor of 3 does not split the zero bias peak, nor makes it disappear. Taken at $V_\mathrm{n-gate} = 2.5$\,V and $V_\mathrm{s-gate} = -2.85$\,V. \textbf{e,} d$I$/d$V$ as a function of $V$, and super gate voltage $V_\mathrm{s-gate}$. The zero bias peak is stable over a finite range in $V_\mathrm{s-gate}$. Taken at $V_\mathrm{n-gate} = 1$\,V and $V_\mathrm{b-gate} = -5.5$\,V.

\end{document}